\documentclass[aps,groupedaddress,superscriptaddress,showkeys,showpacs,twocolumn,nolongbibliography]{revtex4-2}%

\usepackage{amsfonts}
\usepackage{amsmath}
\usepackage{amssymb}
\usepackage{graphicx}
\usepackage{mathcomp}
\usepackage{stmaryrd}
\usepackage[cal=boondox]{mathalfa}

\usepackage{amsmath}

\usepackage{listings}
\usepackage{color}

\definecolor{dkgreen}{rgb}{0,0.6,0}
\definecolor{gray}{rgb}{0.5,0.5,0.5}
\definecolor{mauve}{rgb}{0.58,0,0.82}

\usepackage{mathtools}
\usepackage{braket}
\usepackage{enumitem}

\usepackage[colorlinks=true,allcolors=blue]{hyperref}
\usepackage{graphicx}

\usepackage{savesym}
\usepackage{bm}
\savesymbol{iint}
\usepackage{wasysym}
\restoresymbol{WSS}{iint}

\renewcommand*{\Re}{\operatorname{Re}}
\renewcommand*{\Im}{\operatorname{Im}}

\DeclareMathOperator{\Tr}{Tr}

\begin{document}

\title{Quantum framework for event graphs}

\author{R. P. Erickson}
\email[]{Electronic address: Robert.Erickson2@wellsfargo.com}
\affiliation{Wells Fargo, Phoenix, AZ, USA}

\date{\today}

\begin{abstract}
Graph representations of discrete events provide a natural foundation for machine-learning models of anomaly detection, yet also suggest a deeper quantum description in which graph structure gives rise to interacting quantum degrees of freedom. We develop a quantum framework based on a directed participant graph whose edges represent events connecting pairs of source and destination vertices. A line-graph transformation maps each event to a node of a bidirectional event graph, whose edges inherit relational information from the participant graph. Since event datasets are naturally organized as collections of event records, their raw attributes align directly with nodes of the event graph. A quantum harmonic oscillator (QHO) is assigned to every node of the participant graph, with the collective Hilbert space of these QHOs providing a complete basis for representing quantum states. Every directed edge of the participant graph thereby acquires a Schwinger isospin arising from the two endpoint oscillators. Under the line-graph transformation, event-graph nodes correspond to observable isospins whose interactions through bidirectional edges provide a natural substrate for learning from event datasets, while quantum states associated with underlying participant nodes remain latent and inaccessible to direct observation. Within this framework we formulate a compact U(1) lattice gauge theory (LGT) on the event graph that leads to a Kogut-Susskind Hamiltonian (KSH) in the form of an XY-type spin model governing dynamics of sparse anomalous-event isospins, which we extend, for machine-learning purposes, to a four-state event space with explicit nominal energies and inter-sector couplings. The proposed framework establishes a mathematical foundation for quantum-inspired graph-based anomaly detection and provides a principled bridge between graph learning, LGT, and quantum information.
\end{abstract}


\maketitle

\section{Introduction\label{sec:introduction}}
Graphs provide a natural mathematical framework for representing relational systems, while quantum theory provides a powerful formalism for describing interactions between discrete degrees of freedom. The convergence of these two viewpoints has found expression in a wide range of physical theories, including LGTs \cite{Wilson1974,*Kogut1979}, spin networks \cite{Penrose1971,Baez1996}, quantum walks \cite{Kempe2003,*VenegasAndraca2012}, and tensor-network methods \cite{Orus2014}. In particular, spin networks were originally introduced by Penrose as a combinatorial description of angular momentum couplings \cite{Penrose1971} and later became the foundation of loop quantum gravity (LQG), where graph edges carry irreducible representations of $\mathrm{SU}(2)$ while their intertwiners encode gauge-invariant couplings that describe quantum geometry \cite{RovelliSmolin1995,Baez1996}. An important development was the realization that these spin-network degrees of freedom admit an equivalent representation in terms of pairs of QHOs through the Schwinger construction \cite{osti_4389568,*Sakurai1985-Schwinger}. 

This bosonic formulation has subsequently evolved into a formalism for describing spin networks in terms of QHOs, establishing deep connections between quantum geometry, quantum information, and non-commutative geometry \cite{Girelli2005,*LivineTambornino2012,*Bianchi2015BosonicLQG}. More broadly, the harmonic-oscillator formulation provides a particularly attractive mathematical framework for graph-based systems because local quantum states reside naturally on graph vertices while observable spin variables emerge from interactions between neighboring vertices. These developments suggest that graph-structured systems arising outside quantum gravity may likewise admit physically motivated quantum representations, providing a principled foundation for quantum-inspired models of relational data.

The present work adopts this mathematical framework in a fundamentally different setting. Rather than interpreting graphs as discrete quantum geometries, we instead consider graphs constructed from discrete events recorded within classical datasets. We first construct a directed participant graph in which event participants correspond to graph vertices and events correspond to directed edges. Because machine-learning datasets are naturally organized as collections of event records, representing events as graph nodes aligns the graph representation directly with the native structure of event datasets. Thus, a subsequent line-graph transformation produces an event graph whose nodes represent events and whose bidirectional edges encode relationships inherited from the underlying participant graph. 

The resulting event graph naturally admits a quantum representation in which latent quantum states reside on vertices of the participant graph while observable Schwinger isospins are associated with nodes of the event graph. This quantum representation allows us to formulate a compact U(1) LGT on the event graph, from which a KSH is constructed that describes interactions between sparse anomalous events, which can be extended to a four-state event space with explicit nominal energies and inter-sector couplings, allowing the model to evolve an event graph without class labels as inputs. The resulting Hamiltonian provides a mathematical foundation for the quantum framework developed in this work. The proposed framework combines a graph transformation, from participant to event graph, with a quantum-mechanical transformation, from local oscillators to edge-localized isospins, yielding an event-centric representation in which both topology and latent participant states contribute to the observable event dynamics.

\section{Theory\label{sec:theory}}

\subsection{The Quantum Graph Framework\label{subsec:framework}}
\begin{figure}[t]
\includegraphics[width=240pt, height=250pt]{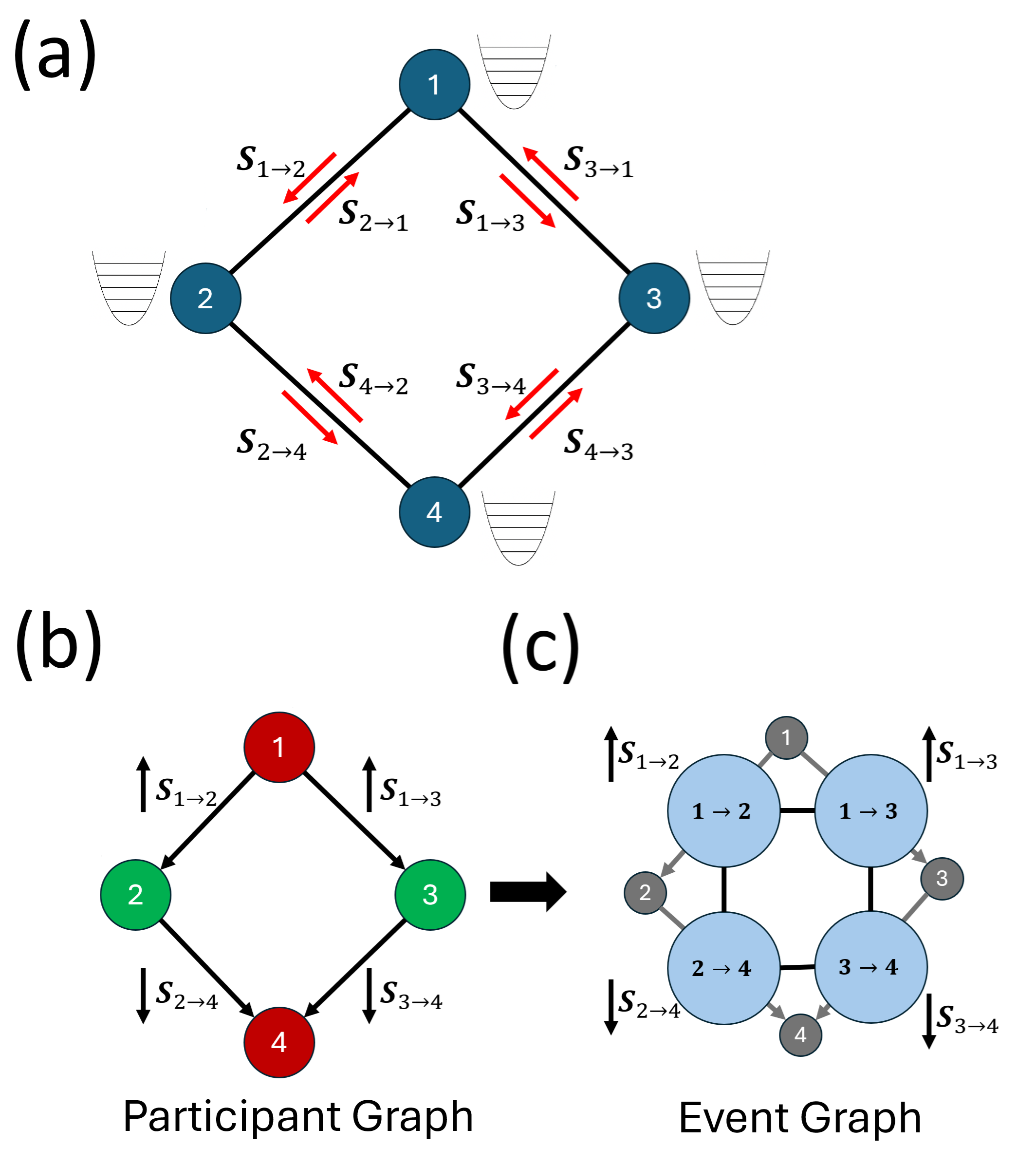}
\caption{\label{fig1}(a) A $4$-node graph with QHO at each participant node and $2$ Schwinger isospins of form $\mathbf{S}_{j\to k}$ and $\mathbf{S}_{k\to j}$. The red arrows indicate edge direction. (b) Simple participant graph with $4$ participant nodes and $2$ event paths $1\to 2\to 4$ and $1\to 3\to 4$. (c) Line-graph transformation of (b) to event graph where event edges become nodes, participant nodes are shaded to indicate latency, and bidirectional edges connect event nodes that have participant nodes in common.}
\end{figure}

The participant graph $G(V,E)$ contains nodes $j, k\in V$ and directed edges $j\to k \in E$, where $j$ is a source and $k$ is a destination. If we place a QHO at each participant node, whether $j$ or $k$, then a Schwinger isospin ${\bf S}_{j\to k}$, a quantum angular momentum operator, can be defined on the directed edge $j\to k \in E$, emergent from the connection between the two QHOs of $j$ and $k$ \cite{osti_4389568,*Sakurai1985-Schwinger}. Figure~\ref{fig1}(a) illustrates placement of QHOs on participant nodes of a simple graph structure, with examples of possible Schwinger isospins bound to directed edges.

Invoking a line-graph transformation on $G(V,E)$, we create the event graph $\tilde{G}(\tilde{V},\tilde{E})$ that contains an isospin at each event node $\ell\in\tilde{V}$. A bidirectional edge $(\ell,\ell^\prime)\in\tilde{E}$ connects event nodes $\ell,\ell^\prime\in\tilde{V}$ if there exists underlying participant nodes in common, thereby defining an interaction between isospins ${\bf S}_\ell$ and ${\bf S}_{\ell^\prime}$. Bidirectional edges of $\tilde{G}(\tilde{V},\tilde{E})$ have an intrinsic binary path-preserving attribute, which is either {\em path-like} (true) or {\em non-path-like} (false). If the attribute is path-like then the bidirectional edge is said to preserve an event path of the original participant graph; otherwise, the attribute is non-path-like. 

Panels (b) and (c) of Fig.~\ref{fig1} illustrate the properties of a line-graph transformation. Panel (b) depicts a simple $4$-participant graph with $4$ events and $2$ event paths, $1\to 2\to 4$ and $1\to 3\to 4$. Panel (c) illustrates the event graph that results from a line-graph transformation, complete with all connecting bidirectional edges. For example, a bidirectional edge connects event nodes $1\to 2$ and $2\to 4$ of (c) because these nodes share underlying participant node $2$. In contrast, events nodes $1\to 2$ and $3\to 4$ are not connected by a bidirectional edge because they do not share any participants. The two bidirectional edges of (c) connecting event node $1\to 2$ with $2\to 4$ and $1\to 3$ with $3\to 4$ are path-like because they preserve paths $1\to 2\to 4$ and $1\to 3\to 4$ of original graph (b), respectively, whereas the two bidirectional edges connecting event node $1\to 2$ with $1\to 3$ and $2\to 4$ with $3\to 4$ are non-path-like because these connections are not associated with event paths of (b).

We present a quantum graph framework by first introducing it on a participant graph. Under a line-graph transformation to the event graph the utility of the framework becomes apparent. Importantly, on the event graph, participant nodes are hidden and their corresponding quantum states are latent, in contrast to isospins at event nodes, which contain accessible information, as gleaned from available event-driven datasets. Our understanding of events occurs at the event-graph level, in the isospins and the way they interact with one another. We next turn to the dynamics of the quantum framework on the event graph, in which anomalous events can be delineated from nominal events.

\subsection{The Relationship Between Latency and Isospin\label{subsec:latency-isospin}}
In principle, the QHO associated with each participant node $j\in V$ possesses an infinite-dimensional Fock space spanned by the basis states $\ket{n}_j$, i.e., $\mathcal{H}^{(\infty)}_j$ where 
\begin{equation}
\mathcal{H}^{(d)}_j=\operatorname{span}\{\ket{n}_j \mid 0\le n<d\} .
\end{equation}
The present work adopts the lowest nontrivial truncation of $d=2$, thereby reducing the local Hilbert space associated with each participant node, $\mathcal{H}^{(2)}_j$, to a two-level system. This choice is motivated not by any fundamental restriction of the proposed framework, but by the resulting correspondence with spin-$1/2$ operators and the Ising-like spin dynamics developed in the following sections. The two-level truncation considered here therefore represents the simplest member of a broader family of event-graph theories. Higher-dimensional local Hilbert spaces of $d>2$ remain fully compatible with the event-graph construction and may instead give rise to effective spin-$S$, bosonic, or other quantum many-body descriptions, depending on the chosen truncation and interaction model. Adopting the two-level truncation ($d=2$), the latent state associated with $j\in V$ becomes
\begin{equation}	\label{eq:latent-state}
\ket{\psi}_j = a_j \ket{0}_j + b_j \ket{1}_j \; ; \quad {|a_j|}^2 + {|b_j|}^2 = 1 .
\end{equation}

Because observable entities of interest are directed events connecting latent participant states, the Schwinger representation provides a natural bridge between local oscillator degrees of freedom and edge-localized angular momentum variables. The resulting isospins inherit information from both endpoint oscillators while furnishing a compact spin description suitable for lattice gauge dynamics on the event graph. Specifically, from the work of Schwinger \cite{osti_4389568,*Sakurai1985-Schwinger}, isospin $\mathbf{S}_{j\to k}$ emerges as a quantum angular momentum operator containing information about the QHO state of both nodes $j,k\in V$. It has Cartesian components $S^\alpha_{j\to k}$, where $\alpha\in\{x,y,z\}$. For the special case $j=k$, we define $\mathbf{S}_{j\to j}=0$, and we note $\mathbf{S}_{j\to k}\ne\mathbf{S}_{k\to j}$ when $j\ne k$. 

In the full Fock space, the isospin is expressible in the second-quantized representation of the two connected QHOs, via annihilation $\hat{a}_j$, $\hat{a}_k$ and creation $\hat{a}_j^\dagger$, $\hat{a}_k^\dagger$ operators, respectively. These operate on the basis states of their respective QHOs, in the manner
\begin{equation}	\label{eq:a-com}
\hat{a}_j \ket{n}_j = \sqrt{n} \ket{n-1}_j , \quad \hat{a}_j^\dagger \ket{n}_j = \sqrt{n+1} \ket{n+1}_j ,
\end{equation}
as illustrated for node $j$. The annihilation and creation operators obey boson commutation relations $[ \hat{a}_j, \hat{a}_k^\dagger ]=\delta_{j,k}$, $[ \hat{a}_j, \hat{a}_k ]=0$, as well as the Hermitian conjugates of these expressions. Specifically, the Cartesian components of isospin are \cite{osti_4389568,*Sakurai1985-Schwinger}
\begin{gather}
\label{eq:Sx} S^x_{j\to k} = \frac{\hbar}{2} \left( \hat{a}_j^\dagger \hat{a}_k + \hat{a}_k^\dagger \hat{a}_j \right) , \\
\label{eq:Sy} S^y_{j\to k} = \frac{\hbar}{2i} \left( \hat{a}_j^\dagger \hat{a}_k - \hat{a}_k^\dagger \hat{a}_j \right) , \\
\label{eq:Sz} S^z_{j\to k} = \frac{\hbar}{2} \left( \hat{a}_j^\dagger \hat{a}_j - \hat{a}_k^\dagger \hat{a}_k \right) , 
\end{gather}
which, by virtue of the commutation relations of annihilation and creation operators, satisfy commutation relations
\begin{equation}
\left[ S^\alpha_{j\to k}, S^\beta_{j\to k} \right] = i\hbar \; \epsilon_{\alpha,\beta,\gamma} \; S^\gamma_{j\to k} ,
\end{equation}
where $\alpha,\beta,\gamma\in\left\{x,y,z\right\}$ and $\epsilon_{\alpha,\beta,\gamma}$ is the Levi-Civita coefficient. 

\begin{figure}
\includegraphics[width=240pt, height=100pt]{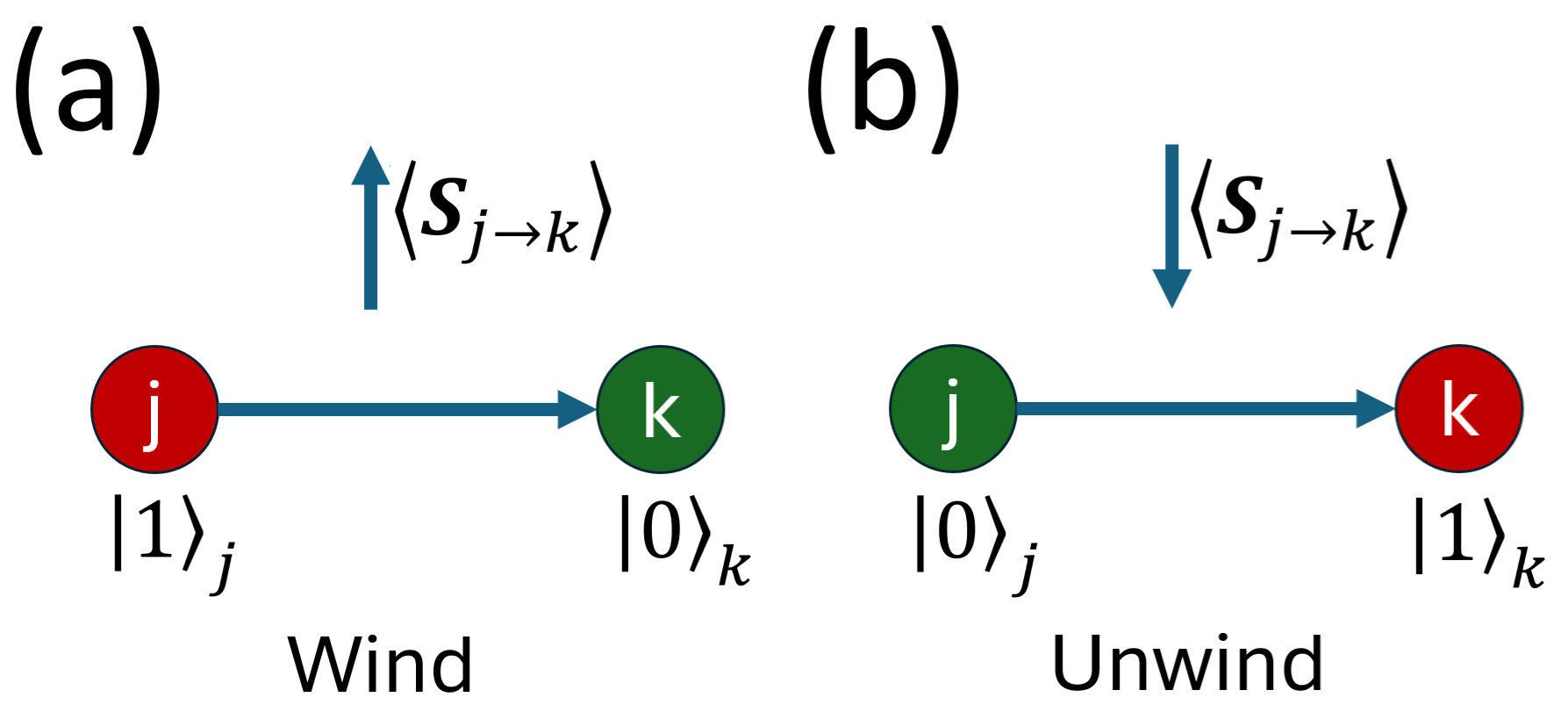}
\caption{\label{fig2}Reference graph examples of binary event archetypes, wind and unwind, expressed via states: (a) $\ket{1}_j \otimes \ket{0}_k$, corresponding to isospin up, and (b) $\ket{0}_j \otimes \ket{1}_k$, corresponding to isospin down.}
\end{figure}

It is useful to consider an example of how information contained in latent states of the form of (\ref{eq:latent-state}) manifests within observable isospins as Ising-like behavior. Suppose that the $N=|V|$ QHOs of the participant graph are initially prepared in a product state
\begin{equation}	\label{eq:psi}
\ket{\psi} = \ket{\psi_1} \otimes \ket{\psi_2} \otimes \cdots \otimes \ket{\psi_N} .
\end{equation}
The expectation value of the isospin associated with directed edge $j\to k\in E$ is then $\braket{{\bf S}_{j\to k}}=\bra{\psi}{\bf S}_{j\to k}\ket{\psi}$. Making use of (\ref{eq:latent-state}) through (\ref{eq:Sz}), we find
\begin{multline}	\label{eq:S-expectation-value}
\braket{{\bf S}_{j\to k}} = \hbar \Big[ \Re \left( a_j^* b_j a_k b_k^* \right) \hat{x} - \Im \left( a_j^* b_j a_k b_k^* \right) \hat{y} \\
+ \frac{1}{2} \left( {\left| b_j \right|}^2 - {\left| b_k \right|}^2 \right) \hat{z} \Big] .
\end{multline}
From (\ref{eq:S-expectation-value}) we see that an isospin can have canted orientation, but the $z$-component has the particular significance of revealing partial information about hidden participant states. This equation explicitly demonstrates that every observable event variable simultaneously encodes information originating from both source and destination participant states.

The expectation value therefore establishes a direct correspondence between latent participant configurations and observable event orientations. Examining (\ref{eq:S-expectation-value}) more closely, when a source participant is in state $\ket{1}_j$ at node $j$ ($b_j=1$) and a destination participant is in state $\ket{0}_k$ at node $k$ ($b_k=0$) then the isospin of (\ref{eq:S-expectation-value}) is in the up state, i.e., $\braket{\mathbf{S}_{j\to k}} = \hbar \hat{z} / 2$, as illustrated in Fig~\ref{fig2}(a). Conversely, when a source participant is in state $\ket{0}_j$ at node $j$ ($b_j=0$) and a destination participant is in state $\ket{1}_k$ at node $k$ ($b_k=1$) then the isospin of (\ref{eq:S-expectation-value}) is in the down state, i.e., $\braket{\mathbf{S}_{j\to k}} = -\hbar \hat{z} / 2$, as illustrated in Fig~\ref{fig2}(b). We refer to these two types of events as purely anomalous events. 

On the other hand, if both participants are in the same state at their respective nodes, whether $b_j=b_k=0$ or $b_j=b_k=1$, then the isospin of (\ref{eq:S-expectation-value}) is zero, i.e., $\braket{\mathbf{S}_{j\to k}} = 0$. We refer to these latter two configurations as purely nominal events. The mixed configurations associated with the two anomalous events are of particular interest because they correspond to directed transitions between distinct local states. As labeled in Fig.~\ref{fig2}, the two anomalous events are basic archetypes to which we assign the designations of {\em wind} and {\em unwind}, which have origin in the latent states of their participants. In contrast, the two nominal states, corresponding to $b_j=b_k=0$ and $b_j=b_k=1$, are basic archetypes to which we assign the designations of {\em background} and {\em foreground}.

Alternatively, the above can be expressed in operator form. Given truncation of Fock spaces to dimension $d=2$, isospin ${\bf S}_{j\to k}$ can be represented by a $4\times 4$ matrix, where we refer to the four corresponding states as (i) the wind state, $\ket{w}_{j\to k} = \ket{1}_j \otimes \ket{0}_k$; (ii) the unwind state, $\ket{u}_{j\to k} = \ket{0}_j \otimes \ket{1}_k$; (iii) the background state, $\ket{b}_{j\to k} = \ket{0}_j \otimes \ket{0}_k$; and (iv) the foreground state, $\ket{f}_{j\to k} = \ket{1}_j \otimes \ket{1}_k$. Again, the wind and unwind states are anomalous states while the background and foreground states are nominal states. From (\ref{eq:a-com}) through (\ref{eq:Sz}), we find
\begin{equation}	\label{eq:pauli-representation}
{\bf S}_{j\to k} \ket{\alpha}_{j\to k} \equiv \left\{
\begin{array}{ccc}
\frac{\hbar}{2} \sigma_{j\to k} \ket{\alpha}_{j\to k} & ; & \alpha \in \{ w , u \} \\
0 & ; & \alpha \in \{ b, f \}
\end{array}
\right. ,
\end{equation}
where $\sigma_{j\to k}$ is a Pauli operator with the usual component definitions, viz.
\begin{gather}
\label{eq:sigma-xy} \sigma^x_{j\to k} = \left(
\begin{array}{cc}
0 & 1 \\
1 & 0
\end{array}
\right) , \quad
\sigma^y_{j\to k} = \left(
\begin{array}{cc}
0 & -i \\
i & 0
\end{array}
\right) , \\
\label{eq:sigma-z} \sigma^z_{j\to k} = \left(
\begin{array}{cc}
1 & 0 \\
0 & -1
\end{array}
\right) .
\end{gather}
Thus, an isospin is essentially a Pauli matrix within a Hilbert subspace of anomalous states. In contrast, via the language of spectroscopy or quantum optics, nominal states are ``optically dark'' since there is no projection of isospin into these states.

If $\tilde{G}_S(\tilde{V}_S,\tilde{E}_S)$ is the subgraph of event graph $\tilde{G}(\tilde{V},\tilde{E})$ having event nodes $\ell\in\tilde{V}_S\subseteq\tilde{V}$ that are expressible as Pauli matrices then $\tilde{G}_S(\tilde{V}_S,\tilde{E}_S)$ is the subgraph of $\tilde{G}(\tilde{V},\tilde{E})$ containing all anomalous events. The Hilbert space of anomalous events can be expressed as
\begin{equation}	\label{eq:hilbert-A}
\mathcal{H}_A = \bigotimes_{\ell\in\tilde{V}_S} \mathcal{H}^{(A)}_\ell \; ; \quad
\mathcal{H}^{(A)}_\ell = \operatorname{span}\{\ket{w}_\ell,\ket{u}_\ell \} ,
\end{equation}
while that of nominal events is
\begin{equation}	\label{eq:hilbert-B}
\mathcal{H}_B = \bigotimes_{\ell\notin\tilde{V}_S} \mathcal{H}^{(B)}_\ell \; ; \quad 
\mathcal{H}^{(B)}_\ell = \operatorname{span}\{\ket{b}_\ell,\ket{f}_\ell \} .
\end{equation}
Treated as an open system, the full Hilbert space is always
\begin{equation}	\label{eq:hilbert-full}
\mathcal{H} = \bigotimes_{\ell\in\tilde{V}} \mathcal{H}_\ell \; ; \quad
\mathcal{H}_\ell = \operatorname{span}\{ \ket{w}_\ell, \ket{u}_\ell, \ket{b}_\ell, \ket{f}_\ell \} ,
\end{equation}
but the partition $\mathcal{H}_A\otimes\mathcal{H}_B$ can evolve in time, or as a function of other time-equivalent external parameters, which requires an account of dynamics.

\subsection{The Quantum Dynamics\label{subsec:dynamics}}
The effective Hamiltonian describing the dynamics of Pauli spins $\{ {\bf\sigma}_\ell \mid \ell\in\tilde{V}_S \}$ is constructed via the Kogut-Susskind formulation \cite{Kogut1975}, retaining those terms associated with the degrees of freedom represented by the event graph. In the present construction, the event-graph links mediate interactions between event-node degrees of freedom but are not assigned independent gauge-field degrees of freedom. Accordingly, the electric- and magnetic-field self-energy terms of the conventional KSH are absent from the effective Hamiltonian.

Furthermore, the local Hilbert-space truncation of $d=2$ reduces the matter sector to Pauli degrees of freedom, as in (\ref{eq:pauli-representation}), while nonlinear contributions merely shift the reference energy and may be removed. Under these assumptions, the resulting effective dynamics take the form of an XY-type spin Hamiltonian, with contributions from energy levels and on-site mixing merged to form an effective Zeeman term. In this way, the KSH of model subgraph $\tilde{G}_S(\tilde{V}_S,\tilde{E}_S)$ is
\begin{equation}	\label{eq:hamiltonian}
H[\sigma] = -\kappa \sum_{(\ell,\ell^\prime)\in \tilde{E}_S} \left( \sigma^+_\ell \tilde{U}_{\ell,\ell^\prime} \sigma^-_{\ell^\prime} + \text{h.c.} \right) - \sum_{\ell\in \tilde{V}_S} \vec{h}_\ell \cdot {\bf\sigma}_\ell ,
\end{equation}
where the hopping term, proportional to coefficient $\kappa$, is a sum over $(\ell,\ell^\prime)\in \tilde{E}_S$. Here, $\sigma^\pm_\ell$ are Pauli ladder operators; ${\bf\sigma}_\ell$ is a vector of  Pauli Cartesian matrices, as in (\ref{eq:sigma-xy}) and (\ref{eq:sigma-z}); $\tilde{U}_{\ell,\ell^\prime}$ is a unitary link operator defined on $(\ell,\ell^\prime)\in \tilde{E}_S$; and $\vec{h}_\ell$ is a local Zeeman field at $\ell\in\tilde{V}_S$. 

Given our earlier note of Ising-like behavior in (\ref{eq:S-expectation-value}), we simplify $H[\sigma]$ by approximating the Zeeman field as $\vec{h}_\ell=h_\ell \hat{z}$. The hopping term of (\ref{eq:hamiltonian}) then permits a phase-valued interaction between neighboring anomalous-event spins. Following the standard LGT construction of Kogut-Susskind \cite{Kogut1975}, the simplest continuous gauge structure associated with such link phases is a compact U(1) gauge field residing on the edges $(\ell,\ell^\prime)\in\tilde{E}_S$. We therefore represent the link operator explicitly as 
\begin{equation}
\tilde{U}_{\ell,\ell^\prime}=\exp{( i \phi_{\ell,\ell^\prime})}\in U(1) ,
\end{equation}
where $0\le\phi_{\ell,\ell^\prime}<2\pi$. The scalar link phase $\phi_{\ell,\ell^\prime}$ parameterizes the relative phase accumulated during interaction between neighboring anomalous-event spins, analogous to the role played by link variables in conventional LGTs. The present work does not require a re-derivation of the underlying gauge formalism; rather, (\ref{eq:hamiltonian}) with $\vec{h}_\ell=h_\ell \hat{z}$ is interpreted within the standard compact-U(1) lattice-gauge framework, from which the effective event-graph Hamiltonian follows. Absorbing the link phase into a complex exchange coefficient
\begin{equation}
J_{\ell,\ell^\prime}=\kappa\exp{(i\phi_{\ell,\ell^\prime})}/2=J_{\ell^\prime,\ell}^* ,
\end{equation}
we arrive at
\begin{equation}	\label{eq:hamiltonian-3}
H[\sigma] = -4 \sum_{(\ell,\ell^\prime)\in \tilde{E}_S} J_{\ell,\ell^\prime} \sigma^+_\ell \sigma^-_{\ell^\prime} - \sum_{\ell\in \tilde{V}_S} h_\ell \sigma^z_\ell .
\end{equation}

Equation~(\ref{eq:hamiltonian-3}) describes the spin-wave dynamics of the anomalous Pauli spins $\{ {\bf \sigma}_\ell \mid \ell\in\tilde{V}_S \}$. The Zeeman field $\{h_\ell \mid \ell\in\tilde{V}_S\}$ of (\ref{eq:hamiltonian-3}) defines the anomaly-event structure, while the hopping terms of $J_{\ell,\ell^\prime}$ introduce a spin-flip susceptibility to disorder. How this anomalous-sector description of dynamics might be applied in a machine-learning context is the subject of the discussion section. Specifically, in the discussion section we propose extending $H[\sigma]$ to include the nominal sector and inter-coupling between anomalous and nominal sectors. As we shall see, this introduces a machine-learning model that can evolve an event graph without class labels as inputs, allowing boundaries between anomalous and nominal events to be learned.

\section{Discussion\label{sec:discussion}}
While the quantum framework developed above is independent of any specific machine-learning architecture, it is instructive to consider how its constituent structures may be incorporated into a graph-transformer encoder. In particular, a multi-head quantum graph transformer, with its attention and message-passing operations, provides an encoder architecture that aligns well to the quantum graph framework, making it an ideal focus for discussion of the utility of our theory \cite{vaswani2017attention,Ying2021}. The following discussion therefore presents one possible realization of the framework rather than a unique experimentally validated implementation.

In the multi-head quantum-graph-transformer context, a density operator of the event graph can be defined that propagates with attention-head index and hidden layer. To permit application of the model to unlabeled event graphs, the anomalous-event Hamiltonian derived in (\ref{eq:hamiltonian-3}) is extended from the anomalous Hilbert subspace to the complete event-state Hilbert space of (\ref{eq:hilbert-full}). The resulting full Hamiltonian retains the KSH of (\ref{eq:hamiltonian-3}) within the anomalous sector, while introducing nominal-sector and anomalous-nominal coupling terms. Its coefficients are determined from the event-node and bidirectional-edge attributes through learned, label-independent parameterizations. 

The resulting Hamiltonian is applied to a soft spectral filter that weighs the similarity of attention and provides the generator of quantum-state evolution. Though outside the scope of the present treatment, specific machine-learning models, with their unique embedding strategies and specific initializations for the type of event they represent, can leverage the prescribed architecture. In what follows we provide a basic outline of encoder architecture for the quantum graph transformer applied to our framework.

\subsection{\label{subsec:density-operator}The Representation of the Density Operator}

\subsubsection{\label{subsubsec:head-encoding}Parallel vs. Sequential Multi-Head Encoding}
Attention-head dynamics of the quantum graph transformer can be realized through either {\em parallel} or {\em sequential} multi-head encoding. Parallel multi-head encoding is the conventional approach adopted by classical multi-head transformer architectures \cite{vaswani2017attention}, while sequential multi-head encoding involves successive head-specific unitary operations performed on the density operator during message passing. Sequential multi-head encoding is of particular interest to the quantum graph transformer design because it potentially incorporates quantum interference and entanglement effects into message passing, thereby offering a quantum-specific form of expressivity. For this reason we focus our presentation to sequential multi-head encoding, leaving the parallel case to the reader. For the sequential multi-head case, we express the event-graph density operator of the transformer as
\begin{equation}	\label{eq:discussion-density}
\rho_{k,m} \in \mathcal D(\mathcal H) \; ; \qquad 0 \le k \le L \; ; \quad 1 \le m \le M , 
\end{equation}
where $L$ is the number of hidden layers, $M$ is the number of attention heads, and the density space is
\begin{equation}	\label{eq:density-space}
\mathcal D(\mathcal{H}) =
\left\{
\rho \mid
\rho=\rho^\dagger,\;
\rho\succeq0,\;
\operatorname{Tr}(\rho)=1
\right\} .
\end{equation}

\subsubsection{\label{subsubsec:approximation}The Transformer Approximation}
For an event graph containing a large number of nodes, direct representation of the formal many-event density operator is generally intractable because the dimension of the tensor-product Hilbert space grows exponentially with number of events. Therefore, practical graph transformers employ a factorized reduced-state approximation in which a local density operator
\begin{equation}	\label{eq:local-density-operator}
\rho^{(k,m)}_\ell\in\mathcal D(\mathcal{H}_\ell) ,
\end{equation}
is maintained for each event node and updated through interactions with neighboring event nodes. Each local Hilbert space retains the complete four-state structure of $\mathcal{H}_\ell$ of (\ref{eq:hilbert-full}), while local density operators approximate corresponding one-event marginals of the formal global state via
\begin{equation}
\rho^{(k,m)}_\ell \cong \Tr_{\ne\ell} \rho_{k,m} .
\end{equation}
Thus, approximation concerns factorization of the many-event density operator into locally maintained one-event reduced states; the four-state Hilbert space $\mathcal{H}_\ell$ of each event is retained in full.

\subsection{The Transformer Hamiltonian\label{subsec:transformer-hamiltonian}}
The Hamiltonian of (\ref{eq:hamiltonian-3}) describes the dynamics of Pauli spins $\{ {\bf\sigma}_\ell \mid \ell\in\tilde{V}_S \}$ of the anomalous-event subspace $\mathcal{H}_A$ of (\ref{eq:hilbert-A}) corresponding to the anomalous-event subgraph $\tilde{G}_S(\tilde{V}_S,\tilde{E}_S)$. In a supervised machine-learning model this information is available during training but not inference. We therefore must extend (\ref{eq:hamiltonian-3}) to an effective transformer Hamiltonian defined on the complete event-state Hilbert space $\mathcal{H}$ of (\ref{eq:hilbert-full}), which includes both the anomalous and nominal sectors, as well as the inter-coupling between them.

The first step is to define $H^{(A)}_{k,m}$, analogous to (\ref{eq:hamiltonian-3}), in terms of winding and unwinding basis states. We write this Hamiltonian as
\begin{multline}	\label{eq:hamiltonian-A}
H^{(A)}_{k,m} = -4 \sum_{(\ell,\ell')\in\widetilde{E}} J_{\ell,\ell'}^{(k,m)} \left( \ket{w}_\ell \bra{u} \right) \otimes \left( \ket{u}_{\ell^\prime} \bra{w}  \right) \\
- \sum_{\ell\in\widetilde{V}} h_\ell^{(k,m)} \left( \ket{w}_\ell \bra{w} - \ket{u}_\ell \bra{u} \right) ,
\end{multline}
where learned exchange and Zeeman coefficients are $J_{\ell,\ell'}^{(k,m)}={J_{\ell^\prime,\ell}^{(k,m)}}^*$ and $h_\ell^{(k,m)}$, respectively. The second step is to define a Hamiltonian, $H^{(B)}_{k,m}$, in terms of background and foreground basis states. A minimal implementation is one diagonal in these states, viz. 
\begin{equation}	\label{eq:hamiltonian-B}
H^{(B)}_{k,m} = \sum_{\ell\in\widetilde{V}} \left( \epsilon^{(k,m)}_{b,\ell} \ket{b}_\ell \bra{b} + \epsilon^{(k,m)}_{f,\ell} \ket{f}_\ell \bra{f} \right) ,
\end{equation}
where learned energy coefficients are $\epsilon^{(k,m)}_{b,\ell}$ and $\epsilon^{(k,m)}_{f,\ell}$, respectively. The simplicity of the choice of (\ref{eq:hamiltonian-B}) does not introduce nominal-sector hopping dynamics and allows the spectral position of nominal states to depend on event attributes. 

As a last step, we introduce $V^{(AB)}_{k,m}$, which couples nominal and anomalous sectors through single-event transitions between all four basis states. The simplest definition in this case is to allow transitions from nominal to anomalous states in the manner
\begin{multline}	\label{eq:hamiltonian-AB}
V^{(AB)}_{k,m} = \sum_{\ell\in\widetilde{V}} \Big( g^{(k,m)}_{w,b;\ell} \ket{w}_\ell \bra{b}+ g^{(k,m)}_{u,b;\ell} \ket{u}_\ell \bra{b} \\
+ g^{(k,m)}_{w,f;\ell} \ket{w}_\ell \bra{f} + g^{(k,m)}_{u,f;\ell} \ket{u}_\ell \bra{f} 
\Big) ,
\end{multline}
where learned coefficients are $g^{(k,m)}_{w,b;\ell}$, $g^{(k,m)}_{u,b;\ell}$, $g^{(k,m)}_{w,f;\ell}$, and $g^{(k,m)}_{u,f;\ell}$. The adjoint ${V^{(AB)}_{k,m}}^\dagger$ defines reverse transitions. Thus, the full transformer Hamiltonian is defined as
\begin{equation}	\label{eq:transformer-hamiltonian}
H^{(\mathrm{full})}_{k,m} = H^{(A)}_{k,m} + H^{(B)}_{k,m} + V^{(AB)}_{k,m} + V^{(AB)\dagger}_{k,m}.
\end{equation}
The first two terms on the right of \eqref{eq:transformer-hamiltonian} describe dynamics within the local anomalous and nominal sectors, respectively, while the last two couple these sectors at each event. The anomalous exchange terms of $H^{(A)}_{k,m}$ act jointly on pairs of connected events.

In summary, we have Hamiltonian coefficients
\begin{multline}	\label{eq:learned-coefficients}
J_{\ell,\ell'}^{(k,m)}, h_\ell^{(k,m)}, \epsilon^{(k,m)}_{b,\ell}, \epsilon^{(k,m)}_{f,\ell}, \\
g^{(k,m)}_{w,b;\ell}, g^{(k,m)}_{u,b;\ell}, g^{(k,m)}_{w,f;\ell}, g^{(k,m)}_{u,f;\ell}
\end{multline}
obtained from event-node and bidirectional-edge attributes through shared learned parameterizations. Consequently, $H^{(\textrm{full})}_{k,m}$ can be constructed for an unlabeled event graph without a priori knowledge of which event nodes are anomalous or nominal. Supervised class labels are used in the training objective rather than in the construction of the Hamiltonian itself. Consistent with our earlier theoretical underpinning, projecting $H^{(\mathrm{full})}_{k,m}$ onto the anomalous subspace recovers the XY spin-model form of \eqref{eq:hamiltonian-3} on the event graph.

\subsection{\label{eq:attention-messaging}Attention and Message Propagation}
Attention can be introduced from a Hamiltonian-derived soft spectral filter acting on the spectrum of $H^{(\textrm{full})}_{k,m}$. Instead of deriving attention from learned attention scores used in transformer architectures \cite{vaswani2017attention} and their graph-transformer extensions \cite{Ying2021}, the present formulation derives attention from the Gibbs-weighted spectrum of $H^{(\textrm{full})}_{k,m}$, corresponding to a normalized matrix exponential familiar from canonical statistical mechanics \cite[Ch.~3]{Pathria1986}. 

Because $H^{(\textrm{full})}_{k,m}$ is constructed from event-node and bidirectional-edge attributes, without requiring class labels, the resulting spectral attention is likewise defined for an unlabeled event graph. Specifically, with $H^{(\textrm{full})}_{k,m}$ arising from (\ref{eq:transformer-hamiltonian}), transformer attention is obtained via the soft spectral filter
\begin{equation}	\label{eq:spectral-function}
A_{k,m} = \text{softmax}_\beta \left( H^{(\textrm{full})}_{k,m} \right) = \frac{e^{\beta H^{(\textrm{full})}_{k,m}}}{\operatorname{Tr}{ \left( e^{\beta H^{(\textrm{full})}_{k,m}} \right) }} ,
\end{equation}
where $\beta$ is an input hyperparameter, nominally set to unity. Operator $A_{k,m}$ senses similarity of connected event nodes through the spectrum of $H^{(\textrm{full})}_{k,m}$, incorporating the learned anomalous, nominal, and inter-sector coupling coefficients of (\ref{eq:learned-coefficients}). 

The spectral attention operator is incorporated into the event dynamics via an attention-weighted effective Hamiltonian
\begin{equation}
H^{(\textrm{eff})}_{k,m} = A_{k,m} H^{(\textrm{full})}_{k,m} A_{k,m} .
\end{equation}
Because $A_{k,m}$ is a spectral filter of $H^{(\textrm{full})}_{k,m}$ then the two operators commute. Moreover, since both $A_{k,m}$ and $H^{(\textrm{full})}_{k,m}$ are Hermitian then $H^{(\textrm{eff})}_{k,m}$ is Hermitian. The spectral attention modifies the relative contribution of eigenmodes of the full transformer Hamiltonian while preserving Hamiltonian evolution, as governed by unitary propagator
\begin{equation}
U_{k,m} = \exp{ \left( - \frac{i}{\hbar} H^{(\textrm{eff})}_{k,m} \Delta t_{k,m} \right) } .
\end{equation}

Propagator $U_{k,m}$ defines messaging as an evolution, with respect to attention head $m$ and hidden layer $k$, dependent on attention-weighted $H^{(\textrm{eff})}_{k,m}$, where input hyperparameter $\Delta t_{k,m}$ specifies the effective evolution interval. Specifically, for sequential head encoding, the density operator propagates in the manner
\begin{equation}	\label{eq:density-propagation}
\rho_{k,m} = U_{k,m} \rho_{k,m-1} U_{k,m}^\dagger ,
\end{equation}
starting from the boundary condition
\begin{equation}
\rho_{k,0} = \left\{
\begin{array}{ccc}
\rho_0 & ; & k = 1 \\
\rho_{k-1,M} & ; & 2 \le k \le L
\end{array}
\right. ,
\end{equation}
where $\rho_0$ is constructed from raw attributes input to the transformer, similar to how learned coefficients of (\ref{eq:learned-coefficients}) are initialized. In this way, the total event state evolves as
\begin{equation}	\label{eq:density-progression}
\rho_0 \to \rho_{1,1} \to \rho_{1,2} \to \cdots \to \rho_{L,M} ,
\end{equation}
such that, on output from the transformer, the final density operator applied to a single readout measurement is $\rho_{L,M}$. The full Hamiltonian encodes event-state interactions, its soft spectral filter determines their attention weighting, and the resulting effective Hamiltonian generates the state evolution.

Nominal background is therefore represented explicitly within $H^{(\textrm{full})}_{k,m}$, rather than through an independently parameterized quantum channel. In particular, $H^{(B)}_{k,m}$ describes the nominal sector and $V^{(AB)}_{k,m}$ couples the nominal and anomalous sectors. Consequently, interactions between these sectors participate directly in spectral attention and subsequent state evolution. Because propagation of (\ref{eq:density-propagation}) is unitary, it also defines a completely positive trace-preserving (CPTP) quantum channel. The quantum-channel representation, derived from $H^{(\textrm{full})}_{k,m}$,  is described in Appendix~\ref{app:quantum-channels}.

\subsection{\label{subsec:training-inference}Training and Inference}
Forward evolution of the transformer described above is independent of event class labels. For a training event graph, $\tilde{G}_\textrm{tr}$, the shared model parameters, collectively denoted by $\theta$, determine from event-node and bidirectional-edge attributes the coefficients of $H^{(\textrm{full})}_{k,m}$, and hence  the spectral attention $A_{k,m}$, effective Hamiltonian $H^{(\textrm{eff})}_{k,m}$, and propagator $U_{k,m}$. Therefore, resulting density operators depend on the observed event graph and the shared model parameters, but not explicitly on the known class labels.

During supervised training, class labels enter only through loss evaluated from transformer readout. Optimization of this loss determines a common set of learned parameters, $\theta^*$, rather than Hamiltonian coefficients associated permanently with individual training-event nodes or edges. Consequently, for a previously unseen event graph, $\tilde{G}_\textrm{inf}$, new coefficients of $H^{(\textrm{full})}_{k,m}$ are generated from attributes of $\tilde{G}_\textrm{inf}$ using fixed learned parameters $\theta^*$. In this way, subsequent spectral-attention and density-evolution calculations of (\ref{eq:spectral-function}) through (\ref{eq:density-progression}) are identical to those used during training and require no a priori identification of anomalous and nominal event subgraphs.

\subsection{\label{subsec:readout}The Transformer Readout}
On output from the transformer, corresponding to $k=L$ and $m=M$, the reduced density operator of event node $\ell\in\tilde{V}$ is
\begin{equation}
\rho^{(\text{out})}_\ell =  \operatorname{Tr}_{\ne \ell}\left( \rho_{L,M} \right) ,
\end{equation}
where the partial trace is taken over all event-node Hilbert spaces other than $\mathcal{H}_\ell$. For realistic event graphs, direct evaluation of the global reduced state is computationally intractable. Consistent with the factorized approximation introduced in the paragraph after (\ref{eq:density-space}), a practical implementation approximates the reduced output state via the locally propagated event-node density operator,
\begin{equation}
\rho^{(\textrm{out})}_\ell = \rho^{(L,M)}_\ell .
\end{equation}
Then classification is obtained directly from decomposition of the local event Hilbert space into anomalous and nominal sectors. 

Specifically, letting orthogonal projection operators be
\begin{equation}	\label{eq:projectors}
P^{(A)}_\ell = \ket{w}_\ell \bra{w} + \ket{u}_\ell \bra{u} , \quad
P^{(B)}_\ell = \ket{b}_\ell \bra{b} + \ket{f}_\ell \bra{f} ,
\end{equation}
such that
\begin{equation}	\label{eq:projector-identities}
P^{(A)}_\ell + P^{(B)}_\ell = I_\ell , \quad
P^{(A)}_\ell P^{(B)}_\ell = 0 ,
\end{equation}
then these define a complete binary measurement of the local event state. Associated with the anomalous and nominal sectors, probabilities 
\begin{equation}	\label{eq:probabilities}
p^{(A)}_\ell = \Tr \left[ P^{(A)}_\ell \rho^{(out)}_\ell \right] , \quad
p^{(B)}_\ell = \Tr \left[ P^{(B)}_\ell \rho^{(out)}_\ell \right] , 
\end{equation}
provide the anomaly score and nominal probability of a given event node $\ell$, respectively. Also, because $\rho^{(\textrm{out})}_\ell$ has unit trace then
\begin{equation}
p^{(A)}_\ell + p^{(B)}_\ell = 1 .
\end{equation}
During supervised binary-cross-entropy training, $p^{(A)}_\ell$ and $p^{(B)}_\ell$ may be used directly, while during inference they provide corresponding class probabilities, without requiring a priori knowledge of event labels.

Importantly, projection measurement does not require anomalous and nominal sectors to decohere during transformer evolution. In particular, off-diagonal components of $\rho^{(\textrm{out})}_\ell$, between subspaces $\mathcal{H}^{(A)}_\ell$ and $\mathcal{H}^{(B)}_\ell$, do not contribute directly to (\ref{eq:probabilities}), but rather such coherence is retained during unitary evolution of the transformer layers, and may therefore influence the final sector probabilities.

\bigskip

Although the proposed framework adopts quantum states, quantum channels, and lattice-gauge dynamics, its computational realization does not require quantum hardware. The resulting quantum graph transformer can be implemented entirely on conventional digital computers using classical numerical linear algebra, with the quantum formalism serving as the underlying mathematical representation rather than as a requirement for quantum circuitry. This permits contemporary graph-learning methods to exploit quantum-mechanical structure while remaining compatible with existing machine-learning software and compute infrastructure.

\section{\label{sec:conclusion}Conclusion}
We have developed a quantum framework for events in which latent participant states give rise to observable Schwinger isospins on an event graph obtained through a line-graph transformation. This construction admits a compact U(1) LGT whose effective KSH describes the dynamics of anomalous event states. The KSH of the anomalous-sector Hamiltonian can be extended to a four-state event space with explicit nominal energies and inter-sector couplings included, allowing for a model that can evolve an event graph without class labels as inputs. The resulting formulation establishes a mathematical foundation for quantum-inspired representations of event-driven relational data.

Although the discussion focused on a quantum graph transformer as one realization of the proposed framework, the underlying theory is independent of any particular learning architecture. It applies equally to event-driven systems exhibiting either adversarial or non-adversarial behavior. The event-graph formalism provides a general foundation for future quantum-inspired graph-learning methods and may be extended to alternative graph models, higher-dimensional local Hilbert spaces ($d>2$), and other event-driven inference problems. Future work will explore the implementation of domain-specific quantum graph transformer architectures derived from the present framework, including applications to financial transaction networks and cybersecurity event streams.

\begin{acknowledgments}
We thank V. Markov and V. Rastunkov for helpful comments and suggestions. The views expressed in this article are those of the authors and do not represent the views of Wells Fargo. This article is for informational purposes only. Nothing contained in this article should be construed as investment advice. Wells Fargo makes no express or implied warranties and expressly disclaims all legal, tax, and accounting implications related to this article.
\end{acknowledgments}

\vspace{0.2in}

\appendix

\section{\label{app:quantum-channels}Quantum Channel Formulation}
The transformer formulation developed above propagates the event-graph density operator at head $m$ and hidden layer $k$ according to (\ref{eq:density-propagation}). Since $H^{(\textrm{full})}_{k,m}$ is Hermitian and $U_{k,m}$ is unitary, then (\ref{eq:density-propagation}) is itself a CPTP quantum channel, with map
\begin{equation}	\label{eq:map-1}
\mathcal{E}_{k,m} (\rho) = K^{(k,m)}_0 \rho {K^{(k,m)}_0}^\dagger ,
\end{equation}
and a single Kraus operator
\begin{equation}
K^{(k,m)}_0 = U_{k,m} \; ; \quad {K^{(k,m)}_0}^\dagger K^{(k,m)}_0 = {U_{k,m}}^\dagger U_{k,m} = I .
\end{equation}
The quantum channel of $\mathcal{E}^{(k,m)}_U$ preserves coherence associated with anomalous and nominal sectors represented explicitly in $H^{(\textrm{full})}_{k,m}$. Equivalent to (\ref{eq:density-propagation}) is the expression
\begin{equation}	\label{eq:density-propagation-2}
\rho_{k,m} = \mathcal{E}_{k,m} (\rho_{k,m-1}) .
\end{equation}

\bibliographystyle{apsrev4-2}
\bibliography{quantum-graph}

\end{document}